\documentstyle[pra,aps,amssymb]{revtex}

\tighten

\begin{document}
\twocolumn[\hsize\textwidth\columnwidth\hsize\csname 
@twocolumnfalse\endcsname
\title{
DECOHERENCE OF ANOMALOUSLY-FLUCTUATING 
STATES OF FINITE MACROSCOPIC SYSTEMS
}
\author{
AKIRA SHIMIZU,\cite{shmz}
TAKAYUKI MIYADERA,\cite{miya}
AND
AKIHISA UKENA
}
\address{
Department of Basic Science, University of Tokyo, 
3-8-1 Komaba, Tokyo 153-8902, Japan
}
\maketitle
\medskip

{\leftskip=5mm \rightskip=5mm
{\small
In quantum systems of a macroscopic size $V$,
such as interacting many particles and quantum computers
with many qubits, there exist pure states such that 
fluctuations of some intensive operator $\hat A$
is anomalously large, 
$
\langle \delta \hat A^2 \rangle = {\cal O}(V^0)
$,
which is much larger than that assumed in thermodynamics,
$\langle \delta A^2 \rangle = {\cal O}(1/V)$.
By making full use of the locality, we show, starting from 
Hamiltonians of macroscopic degrees of freedom, that such states 
decohere at anomalously fast rates when they are weakly perturbed 
from environments.
\par}
\par}
\bigskip
]

\section{Introduction}\label{intro}
\vspace{-3mm}

We consider a quantum system, which extends spatially over 
a {\em macroscopic but finite} volume $V$.
% degrees of freedom $|V|$.
Such a system includes, for example, 
many particles confined in a box of volume $V$, 
and a system composed of $N$ two-level systems (qubits), 
for which $V \propto N \gg 1$.
Such a system, in general, has pure states such that 
fluctuations of some intensive operator,
\begin{equation}
\hat A = \frac{1}{V} \sum_{x \in V} \hat a(x), 
\label{intensiveOp}\end{equation}
where $\hat a(x)$ is an operator 
at point $x$,
is anomalously large;
\begin{equation}
\langle \delta \hat A^2 \rangle = {\cal O}(V^0).
\end{equation}
We call such a pure state an `anomalously fluctuating state' (AFS), 
because $\langle \delta \hat A^2 \rangle$ is much larger than that 
assumed in thermodynamics,
$\langle \delta A^2 \rangle = {\cal O}(1/V)$.

In {\em closed} quantum systems, AFSs appear naturally, 
as explained in section \ref{afs}.
Experimentally, however, it is rare to encounter AFSs.
This apparent contradiction may be explained by the fact that 
real physical systems are not completely closed:
Interactions with environments would destroy AFSs
very quickly.
Effects of environments have been discussed intensively 
in studies of `macroscopic quantum coherence' \cite{AL}, 
and of quantum measurement \cite{Zurek}.
These previous studies assumed
that the principal systems of interest were describable 
by a {\it small number} of collective coordinates,
which interact {\it non-locally} with some environment(s).
However, justification of these assumptions is not clear.
Although such models might be applicable to systems which 
have a non-negligible 
energy gap to excite `internal coordinates' in 
the collective coordinates, 
there are many systems which do not have such an energy gap.
Moreover, 
the results depended strongly on the choices of 
the coordinates and the form of the nonlocal interactions, 
so that general conclusions were hard to draw.
In this work, 
we study the decoherence rates of AFSs and 
normally-fluctuating states (NFSs), 
starting from microscopic Hamiltonians of macroscopic degrees of freedom,
by making full use of the locality:
interactions must be local (Eq.\ (\ref{eqn:int})), 
and macroscopic variables must be 
averages over a macroscopic region (Eq.\ (\ref{intensiveOp})).
To express the locality manifestly, 
we use a local filed theory throughout this work.

\vspace{-3mm}
\section{% Appearance of 
anomalously-fluctuating states 
% of finite systems
}\label{afs}
\vspace{-3mm}

AFSs generally appear in, e.g., 
(i) finite systems which will exhibit symmetry breaking if 
$V$ goes to infinity, 
and (ii) quantum computers with many qubits.

In case (i), 
we can find states (of finite systems)
which approach a symmetry-breaking vacuum as $V \to \infty$.
We call such a state a pure-phase vacuum (PPV).
It has a finite expectation value $\langle \hat M \rangle = {\cal O}(V^0)$ of 
an order parameter $\hat M$, and 
has negligible fluctuations 
$\langle \delta \hat A^2 \rangle = {\cal O}(1/V)$
for {\em any} intensive operator $\hat A$ (including $\hat M$)
\cite{ruelle,haag,HL,SMpre,SMcluster,MS}.
Hence, PPVs are NFSs.
In a mean-field approximation, 
PPVs have the lowest energy.
However, it is known that 
the exact lowest-energy state of a finite system 
(without a symmetry-breaking field)
is generally the {\em symmetric} ground state (SGS), for which 
$\langle \hat M \rangle = 0$ 
\cite{ruelle,haag,HL,SMpre,SMcluster,MS}.
PPVs have a higher (or, in some special cases, equal)
energy than the SGS.
The SGS is composed primarily of a superposition of
PPVs with different values of $\langle \hat M \rangle$
\cite{ruelle,haag,HL,SMpre,SMcluster,MS}.
As a result, it has an anomalously large 
fluctuation of $\hat M$; 
$\langle \delta \hat M^2 \rangle = {\cal O}(V^0)$.
Therefore, if one obtains 
the exact lowest-energy state 
(e.g., by numerical diagonalization) 
in case (i), it is generally 
% the SGS, which is 
an AFS.

In case (ii), 
various states appear in the course of 
a quantum computation.
Some state may be an NFS, for which
$\langle \hat \delta A^2 \rangle = {\cal O}(1/V)$
for {\em any} intensive operator $A$.
This means that correlations between distant qubits are 
weak \cite{MS,US}.
Properties of such states may be possible to emulate by a classical 
system with local interactions, because entanglement is weak.
We therefore conjecture that 
other states -- AFSs -- should
appear in some stage of the computation for a quantum computer to be 
much faster than classical computers.
In fact, we confirmed this conjecture in Shor's algorithm for 
factoring \cite{US}.

We stress that
AFSs are peculiar to {\em quantum} systems of macroscopic but 
{\em finite} sizes, and thus are very interesting.
In (local) classical theories, 
a state such that $\langle \delta A^2 \rangle = {\cal O}(V^0)$
is possible only as a mixed state.
% because superpositions of different states are impossible.
In (local) quantum theory of infinite systems, 
any pure states (including excited states) are NFSs \cite{ruelle,haag},
because all AFSs for a finite $V$ become mixed states in the limit of
$V \to \infty$
\cite{ruelle,haag,SMcluster}.
% (Recall that we require AFSs to be pure.)

% \section{Effects of environments}

\vspace{-3mm}
\section{Interacting many bosons}\label{imb}
\vspace{-3mm}

We first consider 
interacting many bosons confined in a uniform box of volume $V$
with the periodic boundary conditions \cite{SMpre,SMcluster,SMprl}.
Since the Hamiltonian $\hat H$ commutes with the number of bosons $\hat N$, 
there exist simultaneous eigenstates of $\hat N$ and $\hat H$.
We denote the lowest-energy state for a given value of $N$ as 
$|N, {\rm G} \rangle$.
This is the SGS, for which 
$\langle \hat \Psi \rangle = 0$, 
where $\hat \Psi$ denotes the 
intensive order parameter  
$\hat \Psi \equiv (1/V) \sum_{x \in V} \hat \psi(x)$,
and $\hat \psi(x)$ is the boson operator at point $x$ in the box.
On the other hand, $|N, {\rm G} \rangle$ has the long-range order, 
$\langle \hat \psi^\dagger(x) \hat \psi(x') \rangle 
= {\cal O}(V^0)$
for $|x - x'| \sim V^{1/3}$ \cite{SMcluster,SMprl}.
We can easily show that
it has an anomalously-large fluctuation of $\hat \Psi$; 
$ %\begin{equation}
\langle \delta \hat \Psi^\dagger \delta \hat \Psi \rangle = {\cal O}(V^0),
$ %\end{equation}
which shows that $|N, {\rm G} \rangle$ is an AFS.
On the other hand, by superposing $|N, {\rm G} \rangle$ of various 
values of $N$ \cite{converse}, 
we can construct a state $|\alpha, {\rm G} \rangle$, 
for which $\langle \hat \Psi \rangle = {\cal O}(V^0)$
and 
$\langle \delta A^2 \rangle = {\cal O}(1/V)$
for {\em any} intensive operator $A$ \cite{SMcluster}.
Namely, $|\alpha, {\rm G} \rangle$ is a PPV, hence is an NFS.
Although it is not an eigenstate of $\hat H$,
$|\alpha, {\rm G} \rangle$ does not collapse for a 
macroscopic time \cite{SMpre}.

Since the energy of $|N, {\rm G} \rangle$ is
lower (by ${\cal O}(V^0)$) than that of $|\alpha, {\rm G} \rangle$ 
for the same value of $\langle \hat N \rangle$ \cite{SMpre},
there seems to be nothing against the realization of 
$|N, {\rm G} \rangle$ if the system is closed.
However, most real systems are not completely closed, 
and weak perturbations from 
environments can alter the situation dramatically.
In fact, 
effects of interactions with an environment, 
which was assumed to be a huge room that has initially no bosons,
on these states was studied in 
Ref.\ \cite{SMprl}, where 
it was shown that 
$|N, {\rm G} \rangle$ decoheres much faster than 
$|\alpha, {\rm G} \rangle$, as bosons escape from the box 
into the environment.
Namely, the SGS (which is an AFS) is much more fragile 
than the PPV (an NFS), for interacting many bosons in a leaky box.
This may be the first example in which
the SGS and PPVs are identified 
% (where fluctuations of {\em any} intensive 
% operator for PPVs are shown to be negligibly small)
{\em and} the fragility of 
the SGS as well as the robustness of PPVs are shown, 
for a non-trivial interacting many-particle system.

\vspace{-3mm}
\section{General 
% macroscopic 
systems}\label{gs}
\vspace{-3mm}

We next consider a
{\it general} finite system of a {\em large} V, 
interacting with a {\em general} environment E, 
with a {\em local} interaction,
\begin{eqnarray}
\hat H_{\rm int}=
\lambda \sum_{x \in V_{\rm C}} \hat a(x) \otimes  \hat b(x),
\label{eqn:int}
\end{eqnarray}
where $\hat a(x)$ and $\hat b(x)$ are local operators
({\em any} functions of the fields and their conjugate momenta at point $x$) 
of the principal system
and E, respectively,  
{\em at the same point} $x \in V_{\rm C}$.
Here, $V_{\rm C}$ ($\subseteq V$) is a `contact region' 
between the principal system and E.
Since we are interested in the case of weak perturbations from E,
we assume that the coupling constant $\lambda$ ($\geq 0$) is small.

The initial ($t=0$) state, $\rho(0)=|\phi \rangle \langle \phi|$,
of the principal system is either an AFS or an NFS, 
and
the initial state  
of the total system is assumed to be the 
uncorrelated product, 
$\rho_{\rm tot}(0)=\rho(0) \otimes \rho_{\rm E}$,
where $\rho_{\rm E}$ is a time-invariant state of E.
We are interested in the time evolution of the reduced density
operator, 
$\rho(t) \equiv {\rm Tr}_{\rm E}[ \rho_{\rm tot}(t)$],
which generally evolves from the pure state 
$|\phi \rangle \langle \phi|$
into a mixed state.
As a measure of the purity, 
we evaluate the `linear entropy' defined by
$S_{\rm lin}(t) \equiv 1- {\rm Tr}[\rho(t)^2]$,
as a power series of $\lambda^2$,
$S_{\rm lin} = S_{\rm lin}^{(0)} + S_{\rm lin}^{(1)} + \cdots$, 
where $S_{\rm lin}^{(n)} = {\cal O}(\lambda^{2n})$.
We confirmed that this series 
converges \cite{MS}.
Since $S_{\rm lin}^{(0)}=0$, 
this suggests that 
$S_{\rm lin}^{(1)}$ would give the dominant contribution
to $S_{\rm lin}$ under our assumption that $\lambda$ is small.

If $|\phi \rangle$ is translationally invariant \cite{approx}, 
both spatially and temporally, 
we can show that \cite{MS}
\begin{equation}
S_{lin}^{(1)}(\phi,t)
\geq
\frac{\lambda^2}{\hbar^2} 
g_{00}
\langle \phi| \delta \hat A^\dagger \delta \hat A |\phi \rangle
t,
\label{th1}\end{equation}
where 
$\hat A \equiv (1/V)\sum_{x\in V} \hat a(x)$, and
$g$ is a positive matrix defined by
the time correlation of $\hat b$ of E;
\begin{eqnarray}
g_{k_1 k_2} 
\equiv
\frac{1}{2}
\int^{\infty}_{-\infty}ds \langle \hat b_{k_1}^\dagger \hat b_{k_2}(s) \rangle.
\end{eqnarray}
Here, 
$b_k \equiv \sum_{x\in V_{\rm C}} b(x) \mbox{e}^{-ikx}$, where
the sum is not over the entire region of E, but over $V_{\rm C}$.
Since the rhs of Eq.\ (\ref{th1}) is proportional to $t$,
we can interpret it divided by $t$ as
a {\it lower bound of the decoherence rate}, 
which we denote $\gamma$.
% We find that $\tau_1^{\rm dec}$
It is proportional to 
the fluctuation of the intensive operator $\hat A$ composed of 
$\hat a(x)$ which constitutes $H_{\rm int}$.
As discussed in the next section, 
in real physical systems 
many terms would exist in $\hat H_{\rm int}$;
$ %\begin{equation}
\hat H_{\rm int}
=
\hat H_{\rm int}^{[1]} + \hat H_{\rm int}^{[2]} + \cdots,
$ %\end{equation}
where 
$
\hat H_{\rm int}^{[\ell]}
=
\lambda^{[\ell]} \sum_{x \in V_{\rm C}^{[\ell]}} \hat a^{[\ell]}(x)
\otimes  \hat b^{[\ell]}(x).
$
It may be possible to construct a state which is {\em exactly 
robust} (i.e., does not decohere at all) against
{\em one} of $\hat H_{\rm int}^{[\ell]}$'s.
Such a state, however, would be fragile against another 
$\hat H_{\rm int}^{[\ell]}$.
It is therefore important to judge the fragility and robustness 
against {\em all} local interactions.
To $\gamma$ of an NFS, 
all of them give only small contributions, 
except when $g_{00} \geq {\cal O}(V)$,
because
$\hat H_{\rm int}^{[\ell]}$ gives
\begin{equation}
\gamma^{[\ell]} = (\lambda^{[\ell] 2}/\hbar^2) g_{00}^{[\ell]}
\times {\cal O}(1/V),
%\quad \mbox{(NFS)},
\label{g-NFS}\end{equation}
which is small for a large $V$.
To $\gamma$ of an AFS, 
on the other hand, some of $\hat H_{\rm int}^{[\ell]}$'s
can give an anomalously large contribution,
because, 
by definition, 
$
\langle \phi| \delta \hat A^{[\ell] \dagger} 
\delta \hat A^{[\ell]} |\phi \rangle
=
{\cal O}(V^0)
$
for some $\hat A^{[\ell]}$, hence 
\begin{equation}
\gamma^{[\ell]} = (\lambda^{[\ell] 2}/\hbar^2) g_{00}^{[\ell]}
\times {\cal O}(V^0),
% \quad \mbox{(AFS)},
\label{gAFS}\end{equation}
which is larger than Eq.\ (\ref{g-NFS}) by a macroscopic factor 
${\cal O}(V)$.
% This term causes anomalously fast decoherence of AFSs.
% as compared with Eq.\ (\ref{g-NFS}).
%
To see more details,
we estimate the prefactor $g_{00}$ \cite{MS}.
Let $V_{\rm E}^{\rm corr}$ be 
the size of the region in which 
$\int^{\infty}_{-\infty}dt \langle b^{*}(x) b(0,t) \rangle$
is correlated in E.
We can roughly estimate that $g_{00} \propto V_{\rm C}^2$
when $V_{\rm E}^{\rm corr} > V_{\rm C}$, 
whereas
$g_{00} \propto V_{\rm C} V_{\rm E}^{\rm corr}$
when $V_{\rm E}^{\rm corr} < V_{\rm C}$ \cite{MS}.
In either case, the rhs of Eq.\ (\ref{gAFS}) becomes macroscopically
large if the contact region $V_{\rm C}$ is macroscopically large.
Namely, AFSs are fragile.
On the other hand, unlike the case of section \ref{imb},
we cannot draw a general conclusion on the robustness of NFSs
for general systems.

\vspace{-3mm}
\section{Effective theories}
\vspace{-3mm}

Usually, we are only interested in phenomena in some energy range $\Delta E$.
Hence, it is customary to analyze a physical system 
by an effective theory which correctly describes the system only 
in $\Delta E$.
In some cases, 
the degrees of freedom $N$ of the effective theory can become small
even for a macroscopic system
when, e.g., a non-negligible energy gap exists in $\Delta E$
because then the number of quantum states in $\Delta E$ can be small.
% Such a system is essentially a system of small degrees of freedom.
Some SQUID systems are such examples.
We here exclude such systems, and 
concentrate on systems whose $N$ is a macroscopic number,
because otherwise 
the difference between ${\cal O}(1/V)$ and ${\cal O}(V^0)$ 
would be irrelevant (where $V \propto N$).
The effective theory can be constructed from 
an elementary dynamics by an appropriate renormalization 
process.
In this process, in general, 
many interaction terms would be generated in 
the effective interaction $H_{\rm int}$ \cite{Hlocal}.
Hence, it seems rare that 
$H_{\rm int}$ does not have any term 
which takes the form of Eq.\ (\ref{eqn:int}) such that 
$
\langle \delta A^\dagger \delta A \rangle
=
{\cal O}(V^0)
$
for the AFS under consideration.

% In usual analyses of physical systems, 
% one may drop terms with small $\lambda^{[\ell]}$.
% However, for the analyses of decoherence of
% AFSs, such an approximation could be wrong, 
% because, as discussed in the previous section,
% its contribution might be enhanced 
% by a macroscopic factor
% and become relevant, however small $\lambda^{[\ell]}$ is.
If $N$ were small, 
one could drop terms with small $\lambda^{[\ell]}$.
Since $N$ is large, however, 
such an approximation can be wrong, 
because, as discussed in the previous section,
its contribution may be enhanced 
by a macroscopic factor
and become relevant to AFSs, however small $\lambda^{[\ell]}$ is.

\vspace{-3mm}
\section{Implications}
\vspace{-3mm}

We finally discuss implications of the above results,
 for % systems 
(i) and (ii) of section \ref{afs}.

In case (i), our results suggest a new origin of symmetry 
breaking in finite systems \cite{SMprl}.
Although symmetry breaking is usually described as a property of
infinite systems, it is observed in finite systems as well.
Our results suggest that 
although a PPV (which is an NFS) has
a {\em higher} energy than the SGS (an AFS),
the former is realized because the latter is extremely fragile
in environments.
This scenario, may be called `environment-induced symmetry breaking'
after Zurek's `environment-induced superselection rule' \cite{Zurek},
may be the physical origin of symmetry 
breaking in finite systems.

In case (ii), our results 
show that
the decoherence rate can be estimated by fluctuations of 
intensive operators, which depend strongly on the 
number of qubits $N$ and the natures of the states 
of the qubits \cite{ekert}.
This may become a key to the fight against
decoherence in quantum computation.

In both cases, 
we stress that the {\em approximate} robustness 
against {\em all} local interactions (between the principal system
and environments) would be more important than the {\em exact} robustness 
against a particular interaction, because, as discussed in section \ref{gs},
many types of interactions would coexist in real physical systems, 
and the exact robustness against one of them could imply 
fragility to another.
% and the latter would be fragile to one of them.

%%%%%%%%%%%%%%%%%%%%%%%%%%%%%%%%%%%%%%%%%%%%%%%%%%%%%%%%%%%%%%%%%%


\begin{references}
\bibitem{converse}
Conversely, 
$|N, {\rm G} \rangle$ is a superposition of 
$|\alpha, {\rm G} \rangle$
over all values of the angle of $\alpha$.
\bibitem{approx}
This assumption is not satisfied by general AFSs and NFSs.
We here apply Eq.\ (\ref{th1}) to 
some classes of them for which the assumption 
is approximately satisfied.
\bibitem{Hlocal}
We expect that by an appropriate renormalization process 
$H_{\rm tot}$ can be made local in the relevant space-time
scale.
\end{references}
\end{document}